\documentclass[aps,print,prb,twocolumn,superscriptaddress]{revtex4-2}
\usepackage[hidelinks]{hyperref}
\hypersetup{
	colorlinks,
	linkcolor={red},
	citecolor={blue},
	urlcolor={blue}
}

\usepackage{balance}
\usepackage{amssymb}
\usepackage{suffix}
\usepackage{mathtools}
\usepackage[utf8]{inputenc}
\usepackage{booktabs}
\usepackage{cases}
\usepackage[multiple]{footmisc}
\usepackage{dcolumn}
\usepackage{color,soul}
\usepackage{rotating}
\usepackage{perpage}
\usepackage{siunitx}
\usepackage{xcolor}
\usepackage{soul}
\usepackage{amsmath}
\usepackage{tikz}
\usepackage[T1]{fontenc}
\usepackage{etoolbox}
\usepackage{graphics}
\usepackage{siunitx}
\usepackage{float}	
\usepackage{collref}
\usepackage{multirow}
\usepackage{mathtools}
\usepackage{bm}
\usepackage{url}

\usepackage{tikz}
\usepackage{tikz-3dplot}
\usepackage{accents}

\makeatletter
\newcommand{\doublewidetilde}[1]{{%
		\mathpalette\double@widetilde{#1}%
}}
\newcommand{\double@widetilde}[2]{%
	\sbox\z@{$\m@th#1\widetilde{#2}$}%
	\ht\z@=.9\ht\z@
	\widetilde{\box\z@}%
}
\makeatother

\usepackage{etoolbox,lipsum}

\begin{document}

\title{Lattice tuning of charge and spin transport in $\beta_{12}$-borophene nanoribbons}

\author{Masoumeh Davoudiniya}
\affiliation{Department of Physics and Astronomy, Uppsala University, Box 516, 751\,20 Uppsala, Sweden}
\author{Jonas Fransson}
\affiliation{Department of Physics and Astronomy, Uppsala University, Box 516, 751\,20 Uppsala, Sweden}
\author{Biplab Sanyal}
\email{For correspondence: Biplab.Sanyal@physics.uu.se}
\affiliation{Department of Physics and Astronomy, Uppsala University, Box 516, 751\,20 Uppsala, Sweden}

\date{\today}

\begin{abstract}
\small
$\beta_{12}$-borophene nanoribbons (BNRs) exhibit magnetic zigzag edges, while other edge configurations are nonmagnetic. However, when the source, central, and drain regions of a logic device are all composed of zigzag BNRs (ZBNRs), the resulting spin polarization remains weak, unless a high voltage is applied. In this work, we demonstrate that lattice vibrations—introduced for example, via a thermal bath coupled to the central BNR—can enhance spin polarization in ZBNRs. This enhancement manifests as marked changes in the current-voltage characteristics, enabling direct experimental probing. In contrast, nonmagnetic edge configurations exhibit phonon-enhanced charge transport. We employ a tight-binding approach augmented with local electron-phonon interactions described by the Holstein model, and compute the phonon-renormalized Green’s functions and transport currents using the Landauer-Büttiker formalism. The mechanism is supported by analyzing both spinless and spinful electronic dispersions and the corresponding density of states. Compared to the phonon-free edges, structural distortions lead to anisotropic electron-phonon couplings, which significantly modify both charge and spin transport. These results position phonon as an effective tuning parameter for optimizing borophene-based logic devices via engineered edge configurations.
\end{abstract}

\maketitle
{\allowdisplaybreaks

\section{Introduction}\label{intro}
Two-dimensional (2D) quantum materials exhibit distinctive electronic and spintronic properties compared to their bulk counterparts. Various 2D materials, such as Xenes (graphene~\cite{Novoselov2005,Gao2018}, phosphorene~\cite{Chaudhary_2022}, stanene~\cite{Zhao2020}, germanene~\cite{ZANDVLIET202227}, silicene~\cite{Kharadi_2020}, and borophene~\cite{Feng2021}), transition metal dichalcogenides~\cite{Zhao2021}, MXenes~\cite{Naguib}, boron nitride~\cite{Roy}, carbon nitrides~\cite{C7CP02711G}, metal nitrides~\cite{Ashraf2020}, and transition metal oxides~\cite{Lany_2015} have been extensively studied. Among these, Xenes have attracted significant attention due to their Dirac-like electronic dispersion, positioning them as promising candidates for next-generation electronic and energy-storage applications. However, practical implementations of many 2D materials are often limited by factors such as low carrier densities in graphene or modest mobilities in semiconducting transition metal dichalcogenides. 

Borophene, a single-atom-thick boron sheet, exists in several polymorphic phases featuring mixed triangular and hexagonal atomic motifs~\cite{HOU2022107189,D3MA00829K,ARKOTI2024135033,Hou2022}. Unlike graphene's purely hexagonal lattice, which leads to symmetric Dirac cones and semimetallic behavior, borophene's diverse phases ($\alpha$, $\beta_{12}$, $\chi_3$, and striped) exhibit asymmetric Dirac cones and intrinsic metallic behavior, supporting intriguing quantum phenomena and ultrafast electronic responses~\cite{Peng}. The $\beta_{12}$ phase, experimentally realized on substrate, stands out for its remarkable structural stability and pronounced Dirac fermion characteristics, as confirmed by angle-resolved photoemission spectroscopy and theoretical investigations~\cite{Feng2021,PhysRevMaterials.1.021001,Li2024}. Extensive research efforts have further unveiled phenomena such as superconductivity, unconventional magnetism, and anisotropic electron transport within various borophene polymorphs~\cite{oki2024,PhysRevB.96.035425,PhysRevB.98.235430,PhysRevLett.118.096401,PhysRevB.98.134514,PhysRevB.97.125424,PhysRevB.98.054104,PhysRevApplied.21.054016,PhysRevApplied.21.034008}. Additionally, lateral confinement in borophene nanoribbons (BNRs) provides a pathway to engineer size- and edge-dependent electronic structures relevant for spintronic devices. Recent theoretical studies, employing first-principles calculations and nonequilibrium Green’s function methods, demonstrated spin-dependent negative differential resistivity and notable spin-filtering efficiencies in zigzag BNRs at specific biases~\cite{Liu2018,Sun2019}.

\begin{figure*}[t]
\centering
\includegraphics[width=\linewidth]{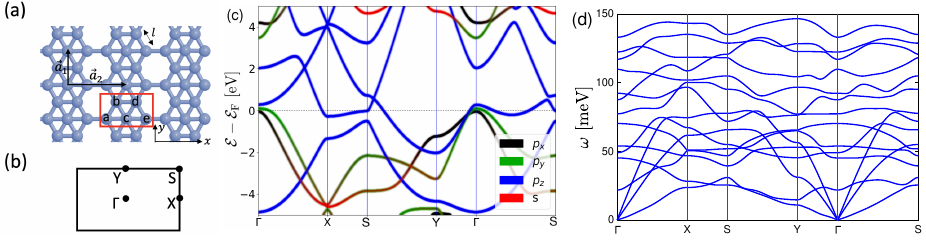}
\caption{(a) Atomic structure of $\beta_{12}$-borophene, showing the unit cell and atomic bonding arrangement. (b) Brillouin zone of the 2D $\beta_{12}$-borophene sheet, indicating the high-symmetry points $\Gamma$, $X$, $Y$, and $S$. (c) Electronic band structure of $\beta_{12}$-borophene with orbital-projected contributions, emphasizing the dominance of the $p_z$ orbital in the vicinity of the Fermi energy. (d) Phonon dispersion, demonstrating the dynamical stability of the $\beta_{12}$-borophene sheet.}
\label{f1}
\end{figure*}

Due to the distinct inherent features of borophene, electron-phonon coupling (EPC) plays a critical role in shaping its electronic and structural properties, significantly impacting carrier lifetimes, optical responses, and thermal stability, distinguishing it from other 2D systems. Recent work has highlighted substantial EPC effects in gated $\beta_{12}$-borophene, where electrode-induced variations led to tunable redshifts or blueshifts in its optical spectrum, a promising feature for optimizing solar cell applications~\cite{phong2024}. Furthermore, first-principles calculations have identified medium-strength EPC in $\beta_{12}$-borophene, predicting superconductivity with transition temperatures around 10 K~\cite{Gao2017}. Studies on hydrogenated borophene derivatives also reported robust EPC, with superconducting transition temperatures predicted up to approximately 20.51 K, significantly extending borophene's potential application space and deepening our understanding of 2D superconductivity~\cite{Chen2023}. These insights highlight EPC's critical role in determining the electronic, thermal, and mechanical responses of borophene, particularly in the technologically promising $\beta_{12}$ phase.

Understanding the role of phonons in charge and spin transport within ultrathin, high-mobility borophene ribbons is also essential for the engineering of next-generation electronic and thermal management applications. Building upon these developments, we pose the question: How do phonons affect the quantum transport properties of $\beta_{12}$-borophene? 

To systematically explore phonon-mediated transport phenomena, we employ a tight-binding Hamiltonian approach and the nonequilibrium Green’s function method combined with the Landauer-Büttiker formalism~\cite{datta_2005} and the Holstein model~\cite{PhysRevB.74.245104,HOLSTEIN1959325,HOLSTEIN1959343}. We investigate both magnetic and non-magnetic BNRs classified by their edge configurations.

The remainder of this paper is structured as follows. Section~\ref{method} presents the computational and theoretical framework, beginning with the computational details, followed by an overview of the geometric structure and the tight-binding Hamiltonian in Subsection~\ref{tb}. The formulation of the EPC strength is introduced in Subsection~\ref{EPC}, followed by the Holstein phonon model in Subsection~\ref{Holstein} and its role in phonon-assisted current in Subsection~\ref{currentph}. Section~\ref{results} investigates the effects of EPC and edge configurations on spin and charge transport in $\beta_{12}$-BNRs, with detailed analyses presented in Subsections~\ref{spinT} and \ref{chargeT}, respectively. Finally, Section~\ref{conclusions} summarizes the key findings of this study.

\section{Methodology}\label{method}

This section details the computational analyses performed, including geometry optimization, electronic structure characterization, and EPC calculations, carried out using the linear combination of atomic orbitals (LCAO) formalism as implemented in the QuantumATK software package~\cite{Smidstrup_2020}. Structural relaxations, electronic band structures, and spin-density distributions were computed using the local density approximation (LDA) with the Perdew–Zunger (PZ) exchange-correlation functional, together with high-quality pseudopotentials from the PseudoDojo library. Brillouin zone integration was performed using a dense Monkhorst–Pack \textit{k}-point grid of $15\times9\times1$, combined with a density mesh cutoff of 90 Hartree, ensuring both numerical accuracy and computational efficiency. Convergence criteria were strictly enforced: self-consistent calculations were terminated when the total energy difference between iterations fell below $10^{-4}$ eV, and geometry optimizations concluded when residual atomic forces were less than $10^{-3}$ eV/Å.
Phonon dynamical matrices and Hamiltonian derivatives were calculated using the finite-difference method with central differences, applying atomic displacements of 0.01 Å. To accurately capture phonon dispersion and EPC properties, these calculations were carried out on a $7\times 5\times 1$ supercell, using a Monkhorst–Pack grid of $5\times 3\times 1$ for momentum-space integration.
For BNR systems, a similar computational protocol was adopted, with adjustments for the quasi-one-dimensional geometry. Specifically, BNR simulations used a \textit{k}-point sampling of $9\times 1\times 1$, a density mesh cutoff of 45 Hartree, and phonon calculations based on $7\times 1\times 1$ supercells.

\subsection{Structural overview and tight-binding model}\label{tb}
Figure~\ref{f1}(a) shows the atomic arrangement of 2D $\beta_{12}$-borophene, where the rectangular red box indicates the primary unit cell. Within this cell, five inequivalent boron atoms, labeled $a$, $b$, $c$, $d$, and $e$, occupy distinct sites. Atoms $a$ and $e$ each bond to four neighbors, atoms $b$ and $d$ bond to five, and atom $c$ bonds to six, resulting in distinct on-site potentials. These differences in bond lengths and angles distinguish $\beta_{12}$-borophene from other boron-based materials and influence its electronic properties, including energy levels and local density of states~\cite{Le2019, Phuong2022}.
Figure~\ref{f1}(b) depicts the rectangular first Brillouin zone of $\beta_{12}$-borophene, identifying high-symmetry points $\Gamma$, $X$, $Y$, and $S$. The calculated lattice parameters are $|\vec{a}_1|=2.94\,\text{\AA}$ and $|\vec{a}_2|=5.09\,\text{\AA}$, consistent with Ref.~\cite{Zhang201615340}.

Figures~\ref{f1}(c) and~\ref{f1}(d) display the DFT electronic band structure and phonon dispersion curves, respectively, for the $\beta_{12}$-borophene sheet. The band structure is shown with orbital-resolved contributions, revealing that the most significant states near the Fermi level arise predominantly from boron $p_z$ orbitals. Notably, the phonon dispersion curves exhibit no imaginary frequencies across the Brillouin zone, indicating that $\beta_{12}$-borophene under study is dynamically stable. From a chemical-bonding perspective, boron atoms in $\beta_{12}$-borophene form a quasi-2D network that exhibits partial $sp^2$-like hybridization. 
Importantly, 2D $\beta_{12}$-borophene can be cleaved along specific crystallographic directions to form $\beta_{12}$-BNRs with various edge configurations. See Sec.\ref{SD-bnr} in the Appendix for further details. The real-space tight-binding Hamiltonian for the $p_z$ orbitals of nearest-neighbor $\beta_{12}$-borophene is given by
\begin{equation} \mathcal{H} = \sum_{i}\varepsilon_{i}\hat{f}^{\dagger}_{i}\hat{f}_{i} + \sum_{\langle i,j \rangle}t_{ij}\hat{f}^{\dagger}_{i}\hat{f}_{j} + \text{H.c.}, \end{equation}
where $\hat{f}_i$ ($\hat{f}_i^\dagger$) is the annihilation (creation) operator on site $i$, $\varepsilon_i$ is the on-site energy, and $t_{ij}$ denotes the hopping amplitude between nearest-neighbor sites $i$ and $j$. The values of these parameters, obtained from DFT calculations~\cite{PhysRevLett.118.096401,PhysRevB.96.035425}, are listed in Table~S1 of the Supplemental Materials (SM)\cite{SM}.

\subsection{Electron-phonon coupling}\label{EPC}
Accurate characterization of EPC is essential to understand the role of lattice vibrations in electronic transport, scattering processes, and low-energy excitations in $\beta_{12}$-BNRs. Phonon frequencies $\omega_{\mathbf{q},\lambda}$ and eigenvectors $\mathbf{u}_{\mathbf{q},\lambda}$, associated with phonon wavevector $\mathbf{q}$ and mode index $\lambda$, are determined by diagonalizing the dynamical matrix $\mathbf{D}(\mathbf{q})$:
\begin{equation}
\mathbf{D}(\mathbf{q})\,\mathbf{u}_{\mathbf{q},\lambda} 
= \omega_{\mathbf{q},\lambda}^2\,\mathbf{u}_{\mathbf{q},\lambda}.
\end{equation}
To explicitly include phonon frequencies and atomic masses, we define the mass-scaled phonon mode vectors:
\begin{equation}
\mathbf{v}_{\mathbf{q},\lambda}(\alpha) 
= \mathbf{u}_{\mathbf{q},\lambda}(\alpha)\,\sqrt{\frac{\hbar}{2M_{\alpha}\,\omega_{\mathbf{q},\lambda}}},
\end{equation}
where $M_{\alpha}$ denotes the mass of atom $\alpha$. Electronic Bloch states $\vert\mathbf{k},j\rangle$ are expanded consistently in a localized-orbital (LCAO) basis set compatible with evaluating gradients of the unperturbed Hamiltonian.

The EPC matrix element for scattering between electronic states $\vert\mathbf{k}, j\rangle$ and $\vert\mathbf{k}+\mathbf{q}, i\rangle$ mediated by phonon mode $\{\mathbf{q},\lambda\}$ is computed as~\cite{Gunst2016,Kaasbjerg2012}:
\begin{equation}
\small
\begin{aligned}
g^{\lambda,\sigma\sigma'}_{ij}(\mathbf{k},\mathbf{q}) =&
\sum_{m,n}
\bigl(c^{\mathbf{k+q}}_{i,\sigma}\bigr)^{*}
      c^{\mathbf{k}}_{j,\sigma'}\;
e^{\,i\left[\mathbf{k}\cdot(\mathbf{R}_{n}-\mathbf{R}_{m})
            -\mathbf{q}\cdot\mathbf{R}_{m}\right]}
\\[4pt]
&\times
\bigl\langle
  i,\sigma,\mathbf{R}_{m}\big|
  \mathbf{v}_{\mathbf{q}\lambda}\cdot\nabla\mathcal{H}(\mathbf{r})
 \big|j,\sigma',\mathbf{R}_{n}
\bigr\rangle,
\end{aligned}
\label{gkq_matrix}
\end{equation}
where $\sigma,\sigma'$ denote spin indices, $c_{j}^{\mathbf{k}}$ are expansion coefficients in the LCAO basis, and the exponential factor explicitly ensures momentum conservation.

In $\beta_{12}$-BNRs, the magnitude of EPC and its impact on the electronic structure sensitively depend on the ribbon's edge geometry and local bonding configurations. To characterize the EPC strength in a more compact manner, we introduce:
\begin{equation}
g_{\mathbf{q}\lambda}^{2}\equiv
\frac{1}{N_{k}N_{\mathrm{orb}}}
\sum_{\mathbf{k}}
\sum_{ij,\sigma\sigma'}
\bigl|
  g^{\lambda,\sigma\sigma'}_{ij}(\mathbf{k},\mathbf{q})
\bigr|^{2},
\label{eq:g_scalar}
\end{equation}
where $N_{k}$ is the total number of $\mathbf{k}$-points used in the Brillouin-zone sampling, and $N_{\mathrm{orb}}$ is the number of orbitals in the basis set. Equation~\eqref{eq:g_scalar} provides a Brillouin-zone averaged measure of EPC strength, explicitly removing momentum $\mathbf{k}$ dependence. However, the electron self-energy, which is typically evaluated using the EPC matrix elements, still implicitly retains $\mathbf{k}$-dependence through the electronic Green's function.% $G^{(0)}(\mathbf{k}-\mathbf{q},\omega)$.

\begin{figure*}[tb]
    \centering
\includegraphics[width=\linewidth]{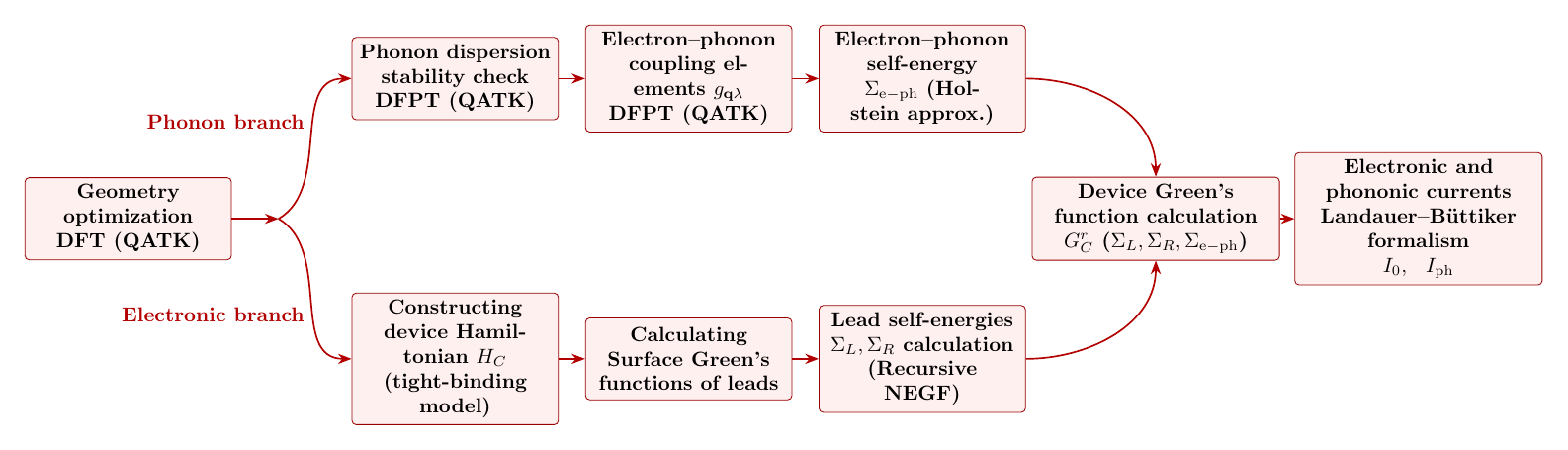}
\caption{Schematic of the multiscale computational framework used in this work. First-principles (ab initio) calculations provide input along two coordinated branches: the upper (phonon) branch yields vibrational modes and EPC elements, while the lower (electronic) branch provides the spin-resolved device Hamiltonian and electrode self-energies. These inputs are integrated into a Green’s function solver, enabling the computation of both elastic and phonon-assisted spin currents within the Landauer–Büttiker formalism.}\label{flowchart}
\end{figure*}

\subsection{Holstein phonon self energy}\label{Holstein}
The effect of electron-phonon interactions on charge transport in $\beta_{12}$-BNRs can be quantitatively described using the Holstein phonon self-energy formalism, which effectively captures local electron-phonon coupling essential for energy and momentum relaxation processes. 
In the absence of electron interactions, phonons are characterized by the non-interacting phonon Green's function:
\begin{equation}
\mathcal{D}^{(0)}_{\mathrm{ph}}(\mathbf{q},i\nu_m)=
\frac{2\omega_{\mathbf{q}\lambda}}
     {(i\nu_m)^{2}-\omega_{\mathbf{q}\lambda}^{2}},
\end{equation}
Note that $i\nu_n$ and $i\nu_m$ label Matsubara frequencies for electrons 
(fermions) and phonons (bosons), respectively~\cite{Mahan2000}. In contrast, $\omega_{\mathbf{q},\lambda}$ is the real-valued phonon frequency 
for mode $\{\mathbf{q},\lambda\}$, obtained by diagonalizing the dynamical matrix.
The EPC modifies the electronic self-energy $\Sigma_{\gamma\eta}(\mathbf{k},i\nu_n)$, evaluated to second-order in perturbation theory as:
\begin{multline}
\Sigma_{\gamma\eta}(\mathbf{k},i\nu_n)=
-\,k_{\mathrm B}T
\sum_{\mathbf{q},\lambda,m}
g_{\mathbf{q}\lambda}^{2}\;
\mathcal{D}^{(0)}_{\mathrm{ph}}(\mathbf{q},i\nu_m)
\\[4pt]
\times\;
G^{(0)}_{\gamma\eta}\!\bigl(\mathbf{k}-\mathbf{q},
                            i\nu_n-i\nu_m\bigr),
\label{eq:sigma}
\end{multline}
where $g_{\mathbf{q}\lambda}$ is the EPC constant defined in Eq.~\eqref{eq:g_scalar}, and $G^{(0)}_{\gamma\eta}(\mathbf{k}-\mathbf{q}, i\nu_n - i\nu_m)$ is the non-interacting electronic Green's function. Here, indices $\gamma$ and $\eta$ explicitly label localized orbital, spin, or sublattice degrees of freedom within the atomic-orbital basis, capturing the detailed internal electronic structure and interactions of the system. The Matsubara frequencies for electrons and phonons are defined as $i\nu_n = i(2n+1)\pi k_B T/\hbar$ and $i\nu_m = i2m\pi k_B T/\hbar$, respectively.

The fully interacting electronic Green’s function incorporates phonon effects through the phonon self‐energy within the NEGF framework. The resulting Green’s function $G_{\gamma\eta}$ is then related to the spectral function $A_{\gamma\eta}(\mathbf{k}, \omega)$ in the usual way~\cite{Mahan2000}:
\begin{equation}
G_{\gamma\eta}(\mathbf{k},i\nu_n) = \int_{-\infty}^{\infty}\frac{d\omega}{2\pi}\,\frac{A_{\gamma\eta}(\mathbf{k},\omega)}{i\nu_n - \omega},
\end{equation}
with the spectral function defined explicitly by:
\begin{equation}
A_{\gamma\eta}(E)=
-2\,\mathrm{Im}\bigl[
  G_{\gamma\eta}(\mathbf{k}, i\nu_n\!\rightarrow\!E+i0^{+})
\bigr].
\end{equation}

For a local Holstein vertex we use the Brillouin-zone–averaged spectrum
\(A_{\gamma\eta}(\omega)=\tfrac{1}{N_{k}}\sum_{\mathbf{k}}
A_{\gamma\eta}(\mathbf{k},\omega)\).

Performing the Matsubara summation explicitly, the self-energy becomes \cite{10.1063/1.4961119}
\begin{multline}
\Sigma_{\gamma\eta}(i\nu_n)=
\frac{1}{2N_{k}}\sum_{\mathbf{q},\lambda}g_{\mathbf{q}\lambda}^{2}
\int_{-\infty}^{\infty}\frac{dE}{2\pi}
\Biggl[
\frac{n_B(\omega_{\mathbf{q}\lambda})+n_F(E)}
     {i\nu_n-E+\omega_{\mathbf{q}\lambda}}\\[4pt]
+
\frac{n_B(\omega_{\mathbf{q}\lambda})+1-n_F(E)}
     {i\nu_n-E-\omega_{\mathbf{q}\lambda}}
\Biggr]A_{\gamma\eta}(E).
\label{eq:SigmaHol}
\end{multline}

Here \(N_{k}\) is the total number of \(\mathbf{k}\)-points used in the Brillouin-zone sampling. \(n_F(E)\) and \(n_B(\omega_{\mathbf q\lambda})\) are the Fermi–Dirac and Bose–Einstein distribution functions, respectively.

\subsection{Phononic effects}\label{currentph}
Phonons significantly affect charge transport by introducing a phonon self-energy into the electronic Green’s function. In a device geometry, the retarded Green’s function for the central region can be written as:
\begin{equation}
G_{C}^{\,r,\sigma}(\omega)=
\Bigl[
 \omega I
 - \mathcal{H}_{C}^{\,\sigma}
 - \Sigma_{L}^{\,r,\sigma}(\omega)
 - \Sigma_{R}^{\,r,\sigma}(\omega)
 - \Sigma_{\mathrm{ph}}^{\,r,\sigma}(\omega)
\Bigr]^{-1},
\label{eq:Gc_ret_spin_sup}
\end{equation}
In this equation, $\sigma$ specifies the spin channel, taking the values spin-up ($\uparrow$) or spin-down ($\downarrow$). $ \mathcal{H}_C $ is the central-region Hamiltonian, $\Sigma_L^r(\omega)$ and $\Sigma_R^r(\omega)$ are the lead self-energies, and $\Sigma_{ph}^r(\omega)$ is the phonon self-energy.

The total charge current $I$, incorporating phonon scattering processes, can be separated into an coherent component $I_0$ and an inelastic phonon-assisted component $I_{ph}$:
\begin{equation}
I^\sigma = I^\sigma_{0} + I^\sigma_{ph}.
\label{eq:CurrentTotal}
\end{equation}
$I_{0}$ measures the current that would flow even if electrons were not scattered by lattice vibrations, whereas $I_{\mathrm{ph}}$ tracks additional carriers that gain (or lose) energy through phonon interactions and thus access conduction channels that are suppressed in purely coherent transport. Including $I_{\mathrm{ph}}$ enhances the overall current and yields a smoother differential conductance profile, illustrating how phonon interactions make a substantial contribution to carrier flow. 
The coherent current $I_{0}$ is given by the Landauer–Büttiker formula in the absence of phonon scattering~\cite{datta_2005}:
\begin{equation}
\footnotesize
I^\sigma_{0} 
= e \int\frac{d\omega}{2\pi}\,\mathrm{Tr}\bigl[\Gamma_L^\sigma(\omega)\,G_C^{r,\sigma}(\omega)\,\Gamma_R^\sigma(\omega)\,G_C^{a,\sigma}(\omega)\bigr]\bigl[f_L(\omega) - f_R(\omega)\bigr],
\label{eq:CurrentI0}
\end{equation}
In this setup, the system consists of a central region coupled to the left and right electrodes, shown in Fig.~\ref{ZZ-w-wo}(a). The Fermi–Dirac distribution functions, $f_{L}$ and $f_{R}$, are associated with the left and right electrodes, respectively, while $\Gamma_{L}$ and $\Gamma_{R}$ represent the coupling (linewidth) matrices of the electrodes.

The phonon-assisted (inelastic) current component $I_{ph}$ arising from EPC is:
\begin{multline}
I^\sigma_{\text{ph}} 
= i\,e\int\frac{d\omega}{2\pi}\,\mathrm{Tr}\Bigl[
\Gamma_L^\sigma(\omega)\,G_C^{r,\sigma}(\omega)\\
\times\left((1 - f_L(\omega))\,\Sigma_{\text{ph}}^{<,\sigma}(\omega) 
+ f_L(\omega)\,\Sigma_{\text{ph}}^{>,\sigma}(\omega)\right)G_C^{a,\sigma}(\omega)
\Bigr],
\label{eq:CurrentIph}
\end{multline}
where $\Sigma_{ph}^{<,>}$ capture phonon emission ($>$) and absorption ($<$) processes, and factors $f_L$ and $1 - f_L$ account for occupied and unoccupied states in the left contact, respectively.
Equations~\eqref{eq:CurrentI0} and~\eqref{eq:CurrentIph} explicitly demonstrate how phonons modify electron transport, broadening electronic states, facilitating energy relaxation, and enabling phonon-assisted tunneling processes. These effects become particularly prominent at elevated temperatures, reflecting increased phonon populations and enhanced electron–phonon scattering.

The computational workflow used in this study is summarized in Fig.~\ref{flowchart}. Following structural optimization and electronic structure calculations performed with QuantumATK, as outlined in earlier sections, the resulting phonon dispersion, EPC elements, and electronic Hamiltonian are incorporated into the device Green’s function formalism. This integrated framework allows for the calculation of both elastic and phonon-assisted currents, as described previously.

\begin{figure*}[t]
\centering
\includegraphics[width=\linewidth]{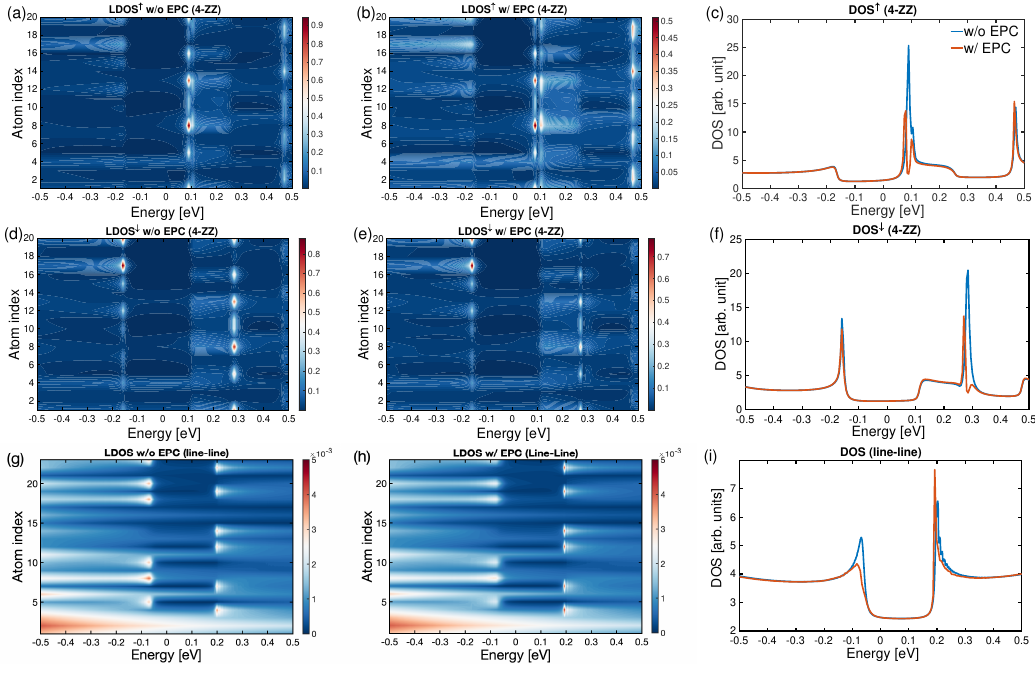}
\caption{Local and total density of states (DOS) for (a–f) zigzag and (g–i) line–line edge configurations of $\beta_{12}$-BNR with a width of 4 unit cells and a length of 4 supercells. Panels (a, b, c) correspond to the spin-up channel of the zigzag $\beta_{12}$-BNR, while panels (d, e, f) represent the spin-down channel. Panels (g, h, i) depict the non-magnetic phase of $\beta_{12}$-BNR with a line–line edge configuration. The left panels (a, d, g) show the LDOS without EPC, whereas the middle panels (b, e, h) include EPC effects. The LDOS plots correspond to the first supercell, with atom indices matching those shown in Fig.~\ref{SD} in the Appendix. The right panels (c, f, i) compare the total DOS with and without EPC. Energy is referenced to the Fermi level, and the color scales indicate LDOS intensity.}
\label{LDOS_4ZZ}
\end{figure*}

\section{Results and Discussion}\label{results}
In previous studies, a dimensionless EPC parameter of about 0.89 was reported for $\beta_{12}$-borophene~\cite{Gao2017}. This number has been reported as a strong strength, as it plays a significant role in $\beta_{12}$-borophene's physical properties. Motivated by these findings, we have explicitly computed the microscopic coupling constants $g_{\mathbf{k},\mathbf{q},\lambda}$ across the Brillouin zone to clarify how phonons affect the charge and spin transport behavior of $\beta_{12}$-borophene. By integrating over the relevant electronic states and phonon modes, we obtain partial contributions from each mode and, from these, a total coupling strength of approximately $0.55\,\text{eV}$ is achieved in the pristine phase. 
Notably, two phonon modes dominate this total strength: an acoustic mode (mode~0) situated near the $\Gamma$ point, with a partial contribution of about $0.2141\,\text{eV}$, and a mid-frequency optical mode (mode~4), contributing around $0.2994\,\text{eV}$. Mode~0 governs long-wavelength scattering processes, while mode~4, involving bond-stretching vibrations, strongly alters electron densities near the Fermi level. Together, they account for more than $90\%$ of the total coupling strength, underscoring their key impact on the transport properties of $\beta_{12}$-borophene. 
The remaining 10\% arises from minor contributions of other phonon modes, which, while less dominant, still play a subtle role in fine-tuning the material’s transport characteristics.

In $\beta_{12}$-BNRs, dimensional confinement and distinct edge configurations further influence EPC characteristics. Structural confinement modifies acoustic phonon dispersion and alters the localization or frequency shifts of optical phonon modes, thereby redistributing their EPC contributions. Edge-induced localization or frequency shifts in phonon modes notably enhance electron scattering. By incorporating mode-resolved EPC data into Holstein self-energy approach, we evaluate phonon-driven spectral renormalizations. Comparative analysis across various ribbon widths and configurations relative to the pristine 2D borophene sheet elucidates critical phonon modes involved in carrier relaxation processes.

In our calculations, we distinguish between two scenarios: a purely coherent regime without phonon effects and a regime with phonons explicitly introduced through Holstein phonons. In the absence of EPC, the lattice is idealized as static, and all carriers propagate coherently without inelastic scattering. Conversely, with EPC, we include site-localized phonons coupled to the electronic degrees of freedom via matrix elements extracted from our DFT calculations. 
The numerical results presented in this work are specifically performed at $T=300$\,K, where the finite thermal phonon population significantly enhances electron--phonon scattering effects compared to the zero-temperature limit.

\subsection{EPC effects on electronic properties of $\beta_{12}$-BNRs}

\begin{figure*}[t]
\centering
\includegraphics[width=\linewidth]{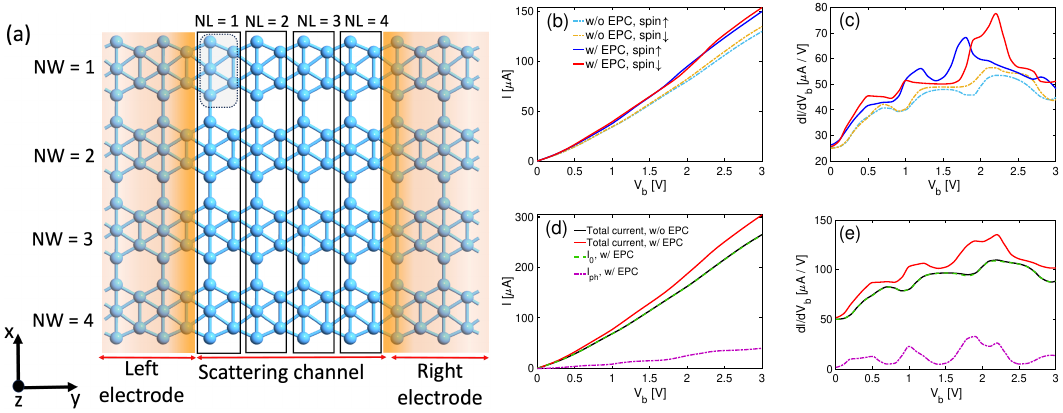}
\caption{(a) Schematic representation of a two-probe model for nonequilibrium Green’s function calculations of a ZZ $\beta_{12}$-BNR. The depicted structure consists of a ribbon that is four fundamental unit cells wide (NW=4) and four supercells long (NL=4). EPC is introduced explicitly within the scattering channel, while the left and right electrodes are indicated at each end. (b)  Current–voltage characteristics comparing spin-up ($\uparrow$) and spin-down ($\downarrow$) conduction channels with (solid lines) and without (dashed lines) EPC. 
(c) The corresponding differential conductance, $dI/dV_{\text{b}}$, derived from the spin-resolved currents in panel (b). (d) Total current ($I_{\text{total}}$) as a function of bias voltage, where the total current with EPC is $I =I_0 + I_{ph}$, explicitly denoting the phonon-induced current contribution (Eqs.~(\ref{eq:CurrentI0}) and (\ref{eq:CurrentIph})). (e) The corresponding total differential conductance illustrating modifications in electronic transport characteristics induced by EPC, derived from panel (d). EPC broadens conduction features, introduces inelastic scattering, and enhances phonon-assisted transport pathways.}
\label{ZZ-w-wo}
\end{figure*}

Although past research has covered borophene’s electronic, magnetic, and phononic properties in various polymorphs~\cite{Gao2025,Xing2025,Kumar2024,Le2019,Sun2019,Liu2018, Davoudiniya2021,Davoudiniya2021_imp}, most of it either focused on pure 2D sheets or used simplified ribbon setups—often missing the detailed impact of EPCs on transport.
In this work, we present a more detailed analysis of $\beta_{12}$-BNRs. We combine a localized-orbital tight-binding approach, Holstein phonons, and a two-probe NEGF method to account for edge-induced magnetism, inelastic phonon scattering, and a wide range of ribbon widths and lengths. 
We explicitly compute spin-dependent EPC matrix elements within polarized scenarios, capturing essential spin-dependent features governing the transport behavior of $\beta_{12}$-BNRs.  
The observation of spin polarization solely in the electronic states of zigzag (ZZ)-terminated $\beta_{12}$-BNRs, as depicted in Fig.\ref{SD}, justifies study of their spin-dependent transport phenomena. 
Our findings reveal spin-dependent renormalization of edge-localized states in ZZ $\beta_{12}$-BNRs, while also showing how phonon scattering affects transport in line–line, AC-AC, and AC–line edge geometries. 
Furthermore, our device-scale simulations offer practical guidance for engineering borophene-based nanoelectronic and spintronic devices by tailoring edge structure and thermal conditions.

Figure~\ref{LDOS_4ZZ} illustrates in detail how edge geometry, spin polarization, and EPC collectively shape the electronic states in a 4-unit-cell-wide $\beta_{12}$-BNR. Panels~(a–f) represent the zigzag-edge configuration, while panels~(g–i) correspond to the line–line edge (non-magnetic) configuration. The left panels (a,d,g) show the layer-resolved local density of states (LDOS) without EPC, middle panels (b,e,h) include EPC, and right panels (c,f,i) compare the total DOS, integrated over the entire scattering region, with (red curves) and without EPC (blue curves). The atomic indices and their corresponding positions for each configuration are explicitly shown in Fig.~\ref{SD}.

For the spin-up channel in the coherent regime [panel~(a)], prominent electronic states clearly appear near $E\approx 0.12\,\mathrm{eV}$. These states display strong LDOS intensities that are distributed rather uniformly across the entire ribbon width, encompassing edge atoms (indices 1 and 20) as well as the interior atomic rows. This uniform distribution demonstrates that these states are not strictly edge-confined but rather spread throughout the full cross-section of the ribbon. Additionally, broader sub-band structures with periodic modulations are observed reflecting quantized transverse electronic modes arising from quantum confinement effects across the ribbon width. Upon introducing EPC [panel~(b)], these electronic states undergo substantial spectral modifications. The sharp peaks around $0.08\,\mathrm{eV}$ broaden by about $50$–$70\,\mathrm{meV}$ and exhibit reduced intensities, indicative of increased electron–phonon-induced inelastic scattering. Importantly, EPC further enhances the spatial homogeneity of LDOS distribution, leading to smoother and less sharply defined intensity modulations across the ribbon interior. Panel~(c) quantitatively captures these EPC-induced spectral changes in the total DOS, illustrating clear peak broadening and diminished intensity, especially around the previously sharp peak at $0.12\,\mathrm{eV}$, along with smoother spectral distributions around the Fermi energy ($E=0$).

In the spin-down channel for the zigzag-edge configuration without EPC [panel~(d)], prominent states emerge distinctly at energies around $E\approx-0.18\,\mathrm{eV}$ and $E\approx0.28\,\mathrm{eV}$, again showing strong intensity localized predominantly but not exclusively at edge atoms. Crucially, these states display substantial LDOS contributions from inner atomic rows as well, reflecting their partially delocalized character. Incorporation of EPC [panel~(e)] significantly alters this spectral distribution. The states around $-0.18\,\mathrm{eV}$ and $0.28\,\mathrm{eV}$ become broadened, with greatly reduced intensity, consistent with electron–phonon-induced lifetime shortening and enhanced inelastic scattering. Furthermore, EPC affects LDOS intensity in interior regions, indicative of increased hybridization between edge-derived and bulk-derived states. The total DOS [panel~(f)] quantitatively confirms these EPC-induced alterations, displaying pronounced broadening and intensity reduction of peaks around $-0.18\,\mathrm{eV}$ and $0.28\,\mathrm{eV}$, alongside smoother spectral distributions, underscoring substantial phonon-mediated spectral reconfiguration.

Furthermore, increasing the borophene nanoribbon width from 5 to 6 unit cells alters the edge-localized states (as shown in Figs.~S2 of the SM~\cite{SM}) increases the number and complexity of edge-localized states, introducing additional phonon modes and scattering pathways that can reinforce or suppress specific features in the DOS. This interrelation between edge geometry, phononic excitations, and spin polarization makes low-dimensional borophene systems sensitive to EPC. Overall, the distinct crystallographic characteristics of ZZ edges in $\beta_{12}$-BNRs centered on the undercoordination of edge atoms—amplify localized states and augment their susceptibility to phonon interactions, rendering EPC a crucial factor in accurately describing the electronic and spin-dependent properties of these systems.

To examine the transport properties of a non-magnetic $\beta_{12}$-BNR, Fig.~\ref{LDOS_4ZZ}(g-i) depicts the LDOS and total DOS for a four-unit-cell-wide ribbon with line-line edges. Calculations are shown (g)~without EPC, (h)~with EPC, and (i) the integrated DOS for the entire ribbon. Only the first supercell is highlighted in the LDOS panels to emphasize where edge-localized states appear (atom indices correspond to Fig.~\ref{SD} in the Appendix.). In the line-line configuration, the uniform edge structure produces sharp peaks at discrete energies, indicative of strong quantum confinement. Edge-localized states, intrinsic to the finite ribbon width, are highly sensitive to electron-phonon interactions, which significantly affect the transport characteristics. For additional configurations (AC-AC and AC-line configurations), see Figs. S3, S4 and S5 of SM~\cite{SM}.

\subsection{Spin transport in magnetic $\beta_{12}$-BNRs with EPC}\label{spinT}

As we observe spin polarization exclusively in $\beta_{12}$-BNRs with ZZ edges, here we investigate how EPC influences the spin transport in these systems. Our findings demonstrate the key role of EPC in modulating the spin-dependent electronic structure of ZZ $\beta_{12}$-BNRs, shedding light on the interplay between edge morphology, spin order, and electron-phonon interactions. %We considered both antiferromagnetic (AFM) and ferromagnetic (FM) configurations, and based on computed total energies, we found the FM configuration (total energy = $-1592.89,\text{eV}$) to be energetically more favorable compared to the AFM configuration (total energy = $-1584.39,\text{eV}$). Therefore, we exclusively focus on the FM configuration throughout this study to thoroughly explore its spin-polarized transport characteristics. 
The spin polarization itself originates from introducing a rigid exchange splitting ($\Delta=0.20\,\text{eV}$~\cite{PhysRevB.96.035425, PhysRevLett.118.096401}), applied uniformly to the spin-degenerate non-magnetic Hamiltonian, as detailed explicitly in Eq.~(S1) of the Supplemental Material. In our transport simulations, the magnetic ZZ $\beta_{12}$-BNR serves as the central scattering region, sandwiched between non-magnetic ZZ $\beta_{12}$-BNR electrodes, ensuring that spin polarization emerges solely from the constructed spin-dependent Hamiltonian of the scattering region.

\begin{figure*}[t]
\centering 
\includegraphics[width=\linewidth]{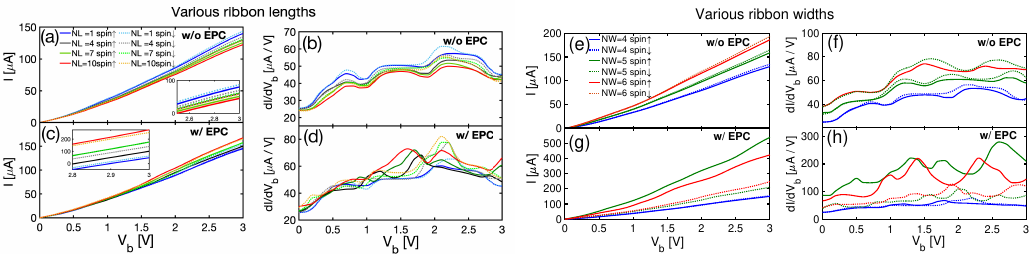}
\caption{Spin-resolved current--voltage characteristics and corresponding differential conductance for a ZZ-$\beta_{12}$-BNR. Panels (a-d) correspond to a fixed ribbon width of four unit cells with varying ribbon lengths ($NL$), while panels (e-h) correspond to a fixed ribbon length of four unit cells with varying ribbon widths ($NW$). Upper panels (a, b, e, f) show results without EPC, and lower panels (c, d, g, h) display data including EPC effects. }
\label{ZZ-diffL}
\end{figure*}

Figure~\ref{ZZ-w-wo}(a) illustrates the two-probe device configuration employed in our NEGF calculations of electron transport through a ZZ $\beta_{12}$-BNRs. The central scattering region consists of four supercells, each composed of four fundamental unit cells along its width; EPC is explicitly included only within this finite region, while the left and right electrodes are modeled as semi-infinite ballistic reservoirs. Panels~(b) and (c) present the spin-resolved current–voltage ($I$–$V_b$) characteristics and their corresponding differential conductances (${\rm d}I/{\rm d}V_b$), respectively, comparing results obtained with EPC (solid lines) and without EPC (dashed lines). Under coherent conditions, panel~(b) shows near-linear current–voltage relationships, reflecting the intrinsic metallicity of these ribbons. At low biases, the currents and differential conductances for both spin channels nearly overlap, indicating negligible spin polarization in this regime. However, as the bias increases, distinct spin polarization emerges and progressively intensifies, a phenomenon clearly discernible from the divergence of the spin-up and spin-down branches in panel~(b) and, more evidently, from the differential conductance curves in panel~(c). Notably, in the presence of EPC, spin polarization appears at significantly lower bias voltages compared to the ballistic case. Furthermore, the inclusion of EPC consistently enhances the total current in both spin channels across the entire bias range. To elucidate the origin of this current enhancement, Fig.~\ref{ZZ-w-wo}(d) decomposes the total EPC-influenced current into coherent ($I_0$) and phonon-assisted ($I_{\mathrm{ph}}$) contributions, revealing that the net current increase predominantly arises from phonon-mediated inelastic processes. Interestingly, the coherent component ($I_0$) remains essentially unaffected by EPC, confirming that elastic transmission channels are only weakly perturbed by electron–phonon interactions. In contrast, the phonon-assisted current ($I_{\mathrm{ph}}$) grows substantially with bias, highlighting the pivotal role of inelastic electron–phonon scattering in activating additional conduction pathways. A detailed analysis of the differential conductance is further provided in Fig.~\ref{LDOS_4ZZ}(e), where the total differential conductance is explicitly resolved into its coherent and phonon-assisted components. Under purely ballistic conditions (without EPC), the differential conductance already displays significant bias dependence, rising gradually from about $50\,\mu$A\,V$^{-1}$ at low bias and exhibiting a broad maximum around $V_b\approx 2$~V, followed by a slight reduction at higher voltages. This non-monotonic behavior directly reflects the intrinsic electronic band structure of the metallic ribbon, influenced by sub-band alignments and van Hove singularities in the density of states near the Fermi energy. Upon introducing EPC, pronounced peaks emerge sharply in the differential conductance around $V_b\approx 1.1$ and $1.7$~V, corresponding precisely to resonant inelastic scattering events mediated by specific phonon modes. At these characteristic biases, electrons efficiently exchange discrete energy quanta with phonons, accessing otherwise inaccessible electronic states and dramatically boosting the conductance. Consequently, the net effect of EPC manifests as a substantial overall increase in differential conductance, dominated by phonon-mediated inelastic contributions.

\begin{figure*}[t]
\centering
\includegraphics[width=\linewidth]{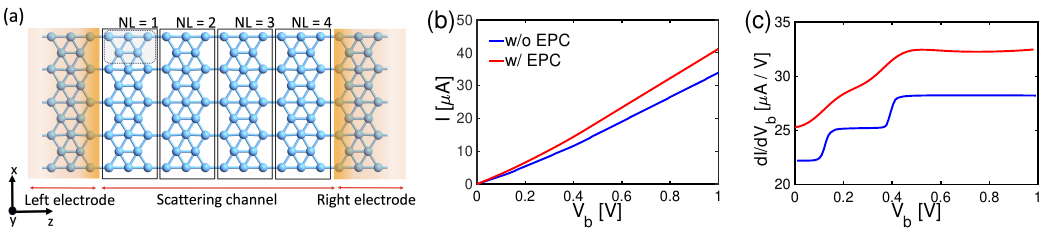}
\caption{(a) Two–probe setup used for the nonequilibrium Green's‑function
    calculations on a 4-unit-cell-wide and 4-supercell-long $\beta_{12}$-BNR with a line–line edge configuration, showing the left and right electrodes along with the scattering channel. The dashed rectangle marks a single repeating unit cell of the $\beta_{12}$-borophene lattice. (b) Current–voltage characteristics computed with and without EPC. (c) Corresponding differential conductance as a function of bias voltage, illustrating sharper conductance features in the absence of EPC. When EPC is introduced, these features become smoother, leading to an overall increase in current.}
\label{LL_diffT}
\end{figure*}

Figure~\ref{ZZ-diffL} compares the current--voltage and differential conductance characteristics of ZZ $\beta_{12}$-BNRs under different transport regimes. Panels (a)--(d) focus on a ribbon with width $NW=4$ and lengths $NL = 1, 4, 7, 10$, whereas panels (e)--(h) examine a fixed length across widths $NW=4, 5, 6$. Solid (dotted) lines indicate spin-up (spin-down) transport, respectively. In all cases, we distinguish between no inelastic scattering and EPC-influenced transport, as well as spin-up and spin-down channels. Under coherent conditions, panel~(a) the current magnitude systematically declines as the ribbon length increases, consistent with exponential decay due to coherent tunneling limitations within the extended scattering region. As highlighted by the inset of panel~(a), this length-induced suppression remains significant even at high biases. The associated differential conductance curves in panel~(b) exhibit pronounced multi-hump features: initially rising near $V_b\sim 0.7$~V, subsequently decreasing around $1$~V, followed by secondary peaks near $1.3$~V and finally leveling off beyond $2.2$~V. These oscillations originate from sub-band crossings and van-Hove singularities in the ribbon's density of states and diminish in amplitude for longer ribbons due to reduced coherent transmission. Moreover, spin-resolved differences become noticeable at biases above $\sim1$~V, with spin-down conductance consistently exceeding spin-up conductance, thereby revealing a clear bias-induced spin polarization amplified with ribbon length. Introducing EPC alters this scenario, as demonstrated in panels~(c) and (d). Panel~(c) reveals that EPC enhances the current across the entire bias range for all ribbon lengths, notably reversing the length dependence observed in panel~(a): the longest ribbon ($NL=10$) now exhibits the highest current. This reversal arises from the dominance of phonon-mediated inelastic transmission, whose contribution scales approximately linearly with increasing ribbon length and thereby surpasses the coherent tunneling limitations. Correspondingly, the differential conductance curves in panel~(d) undergo substantial reshaping: EPC elevates the conductance from approximately $25\,\mu\mathrm{A\,V}^{-1}$ (without EPC) to nearly $30\,\mu\mathrm{A\,V}^{-1}$ for $NL 10$ at zero bias, broadens and smooths the ballistic humps into distinct plateaus, and introduces additional resonant peaks around biases of $1$~V and $1.5$~V. These features reflect bias-dependent activation of specific optical-phonon modes that provide efficient channels for electron energy exchange, thus unlocking otherwise inaccessible conduction pathways.

\begin{figure*}[t]
\includegraphics[width=\linewidth]{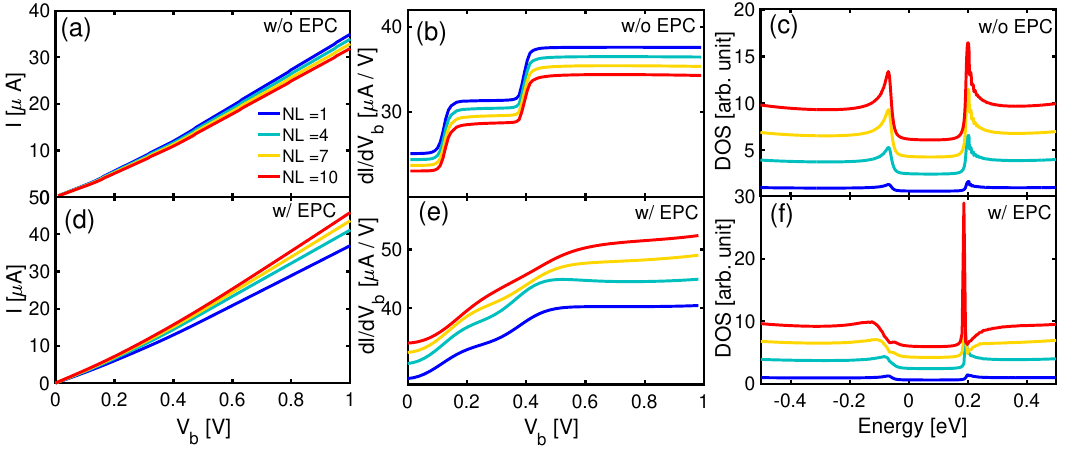}
\caption{Current–voltage characteristics, differential conductance and density of states for a 4-unit-cell-wide $\beta_{12}$-BNR with line–line edge configuration, calculated for different ribbon lengths (NL = 1, 4, 7, 10). Panels (a–c) show results without EPC, highlighting sharper conductance features and distinct DOS peaks due to the absence of scattering. Panels (d–f) illustrate results with EPC effects, demonstrating smoother differential conductance features and DOS profiles. Increasing ribbon length enhances electron–phonon interactions, significantly modifying electronic transport properties.}
\label{LL_diffNL}
\end{figure*}

Figures~\ref{ZZ-diffL}(e)–(h) systematically examine how varying ribbon width ($NW=4$, $5$, $6$) at a fixed length affects the transport characteristics in zigzag-edge $\beta_{12}$-BNRs. Panels~(e) and (f) represent coherent transport (w/o EPC), whereas panels~(g) and (h) include EPC (w/ EPC). In the ballistic limit [panel (e)], the $I$–$V_b$ characteristics are nearly linear, and the magnitude of the current systematically increases with ribbon width, reflecting a greater number of transverse conduction modes available in wider ribbons. Spin-down currents consistently surpass spin-up currents at higher biases ($V_b \gtrsim 2$~V), indicating a clear bias-induced spin polarization arising from width-dependent spin-split states localized at the ribbon edges. Corresponding ballistic differential conductances in panel (f) exhibit distinct multi-peak structures, associated with sub-band crossings and van-Hove singularities in the electronic density of states. These peaks become increasingly pronounced with ribbon width, owing to enhanced transmission through additional conduction pathways. Moreover, spin-down conductance consistently exceeds spin-up across most bias voltages, highlighting intrinsic spin-dependent asymmetry driven by ribbon geometry and edge states.
Upon inclusion of EPC, transport properties dramatically evolve [panels (g) and (h)]. Panel (g) reveals a pronounced EPC-induced enhancement in total current for all ribbon widths. Interestingly, this enhancement is non-monotonic with respect to width, with the intermediate-width ribbon ($NW=5$, green lines) exhibiting the highest total current, surpassing both narrower ($NW=4$) and wider ($NW=6$) ribbons. Such behavior results from competing effects: although wider ribbons offer more scattering sites and conduction channels for phonon-assisted transport, they simultaneously experience enhanced backscattering from increased phonon-induced disorder, thus partially limiting their conductance advantage. Crucially, the spin-dependent current behavior under EPC is notably modified compared to the ballistic regime. In particular, the spin-up current surpasses the spin-down current for the wider ribbons ($NW=5$ and $NW=6$). This remarkable reversal in spin conductance ordering arises due to selective phonon-mediated enhancement of transmission channels predominantly associated with spin-up edge states. Electron–phonon scattering preferentially couples spin-up electrons to previously inaccessible conduction pathways, effectively redistributing spectral weight toward spin-up channels and overcoming their ballistic suppression relative to spin-down channels. This phenomenon is further corroborated by differential conductance curves in panel (h), where EPC-induced resonance peaks with maximum values reaching around $280\,\mu\text{A\,V}^{-1}$ for the $NW=5$ ribbon. These peaks directly result from resonant electron–phonon processes involving specific optical phonon modes, facilitating spin-dependent energy exchange and opening new phonon-mediated conduction channels. Consequently, EPC not only boosts overall conduction efficiency but also provides an effective mechanism for altering spin polarization, particularly by strengthening spin-up conductance beyond spin-down levels in wider ribbons. Thus, these results illustrate how the delicate interplay between electronic structure, phononic interactions, and ribbon geometry influences the transport characteristics, enabling precise control over spin-dependent currents in zigzag-edge BNRs.

\subsection{Charge transport in non-magnetic $\beta_{12}$-BNRs with EPC}\label{chargeT}

To investigate the effects of EPC on charge transport in non-magnetic $\beta_{12}$-BNRs, we performed identical calculations for three different edge terminations: line-line, AC-AC, and AC-line as shown in Fig.~\ref{SD}. The line-line edge configuration consists of straight edges formed by atoms aligned along specific crystallographic directions, creating uniform edge atomic coordination and strongly spatially confined edge states. Our results demonstrate that, while all three configurations exhibit qualitatively similar transport behavior under EPC, the line-line edge configuration consistently yields higher current values. This enhanced conductance is likely attributed to the favorable geometrical arrangement of electronic states at the line-line edges, which effectively promotes charge flow. Consequently, our main discussion focuses on the line-line edge configuration, while the corresponding results for the AC-AC and AC-line configurations are presented in Figs. S3, S4 and S5 of the SM~\cite{SM}.

Figure~\ref{LL_diffT}(a) schematically depicts a $\beta_{12}$-BNR device, 4~unit cells wide and 4~supercells in length, with line-line edges. The dashed rectangle indicates a repeating unit cell in the central scattering region, flanked by left and right electrodes. Figure~\ref{LL_diffT}~(b) presents the current--voltage characteristics with and without EPC, showing that current rises with both EPC and bias. However, in the presence of EPC, a higher phonon population simultaneously introduces inelastic scattering.
Figure~\ref{LL_diffT}~(c) plots the differential conductance, providing information on the underlying transport channels. In the absence of EPC, the nanoribbon exhibits step-like features corresponding to discrete conduction channels set by quantum confinement. Each step signals the onset of a new channel, producing abrupt increases in conductance and a characteristic non-ohmic response. When EPC is included, these steps are smoothed out, and $\mathrm{d}I/\mathrm{d}V$ eventually saturates. Phonon-induced inelastic scattering blurs the quantized energy levels, reducing the prominence of distinct channel activation. In this regime, conduction transitions from being sharply quantized to a smoother, more ohmic-like form.

Finally, Fig.~\ref{LL_diffNL} compares the transport characteristics of a $\beta_{12}$-BNR with line-line edges under two conditions: without EPC [panels~(a--c)] and with EPC [panels~(d--f)]. In the absence of EPC, the differential conductance, $\mathrm{d}I/\mathrm{d}V$, exhibits step-like features tied to discrete conduction channels. Increasing the ribbon length NL intensifies exponential decay of the transmission, reducing the current and narrowing the steps.
Under EPC, phonon-mediated scattering broadens the discrete levels, smooths the steps, and promotes additional transport pathways, allowing longer ribbons to sustain higher currents.

\section{Conclusion}\label{conclusions}
We have systematically investigated how EPC influences the spin and charge transport properties of $\beta_{12}$-BNRs. Using an \textit{ab initio}-based tight-binding model combined with the Holstein approximation and the nonequilibrium Green's function–Landauer formalism, we demonstrate that EPC significantly reshapes both the magnitude and the energy dependence of conduction pathways. Zigzag-edged BNRs enhances the weak inherent spin polarization, whereas other edge configurations such as line–line remain nonmagnetic, while they display higher conductance. EPC consistently enhances the overall current through phonon-assisted inelastic processes, redistributes spectral weights in the density of states, and broadens differential conductance features. These effects underline the importance of lattice degree of freedom and edge-specific electronic structures, positioning EPC and edge engineering as critical strategies for tuning electronic and spintronic performance in borophene-based nanoscale devices.

\section{Acknowledgments}
B.S. acknowledges financial support from Swedish Research Council (grant no. 2022-04309), STINT Mobility Grant for Internationalization (grant no. MG2022-9386) and DST-SPARC, India (Ref. No. SPARC/2019-2020/P1879/SL). The computations were enabled by resources provided by the National Academic Infrastructure for Supercomputing in Sweden (NAISS) at UPPMAX (NAISS 2024/5-258) and at NSC and PDC (NAISS 2023/3-42) partially funded by the Swedish Research Council through grant agreement no. 2022-06725. B.S. also acknowledges the allocation of supercomputing hours granted by the EuroHPC JU Development Access call in LUMI-C supercomputer (grant no. EHPC-DEV-2024D04-071) in Finland.

}

\twocolumngrid
\appendix
{\allowdisplaybreaks

\section{Different configurations of BNRs}\label{SD-bnr}
\begin{figure*}
\centering
\includegraphics[width=\linewidth]{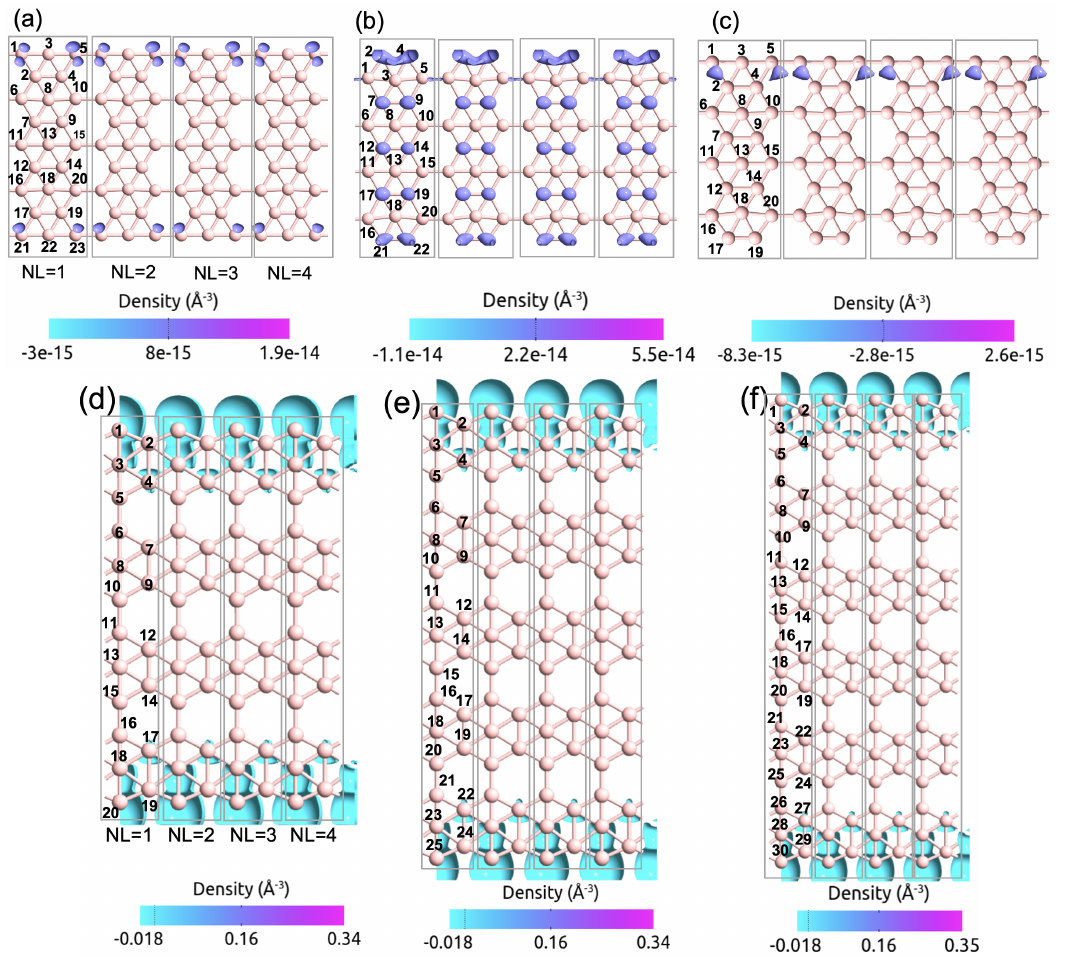}
\caption{Representative ribbon geometries of $\beta_{12}$-12-BNRs featuring various edge terminations and widths. Panels (a)–(c) show (a) line–line, (b) AC–AC, and (c) AC–line terminations, while panels (d)–(f) illustrate ZZ edges for ribbons of widths containing 20, 25, and 30 atoms, respectively. The color scale indicates the spin-density distribution.}
\label{SD}
\end{figure*}
Besides the common Armchair–Armchair (AC–AC) and ZZ edges, line–line and Armchair–Line (AC–line) terminations can occur, significantly affecting electronic transport and device performance. 
Figure~\ref{SD} displays spin density for four edge types: panels~(a) and (b) illustrate line–line and AC–AC terminations, and panel~(c) depicts the AC–line configuration, all exhibiting negligible spin density. Conversely, the ZZ edges shown in panels~(d)-(f), corresponding to ribbon widths of 20, 25, and 30 atoms, respectively, exhibit significant spin-polarized electronic states. 
The calculated magnetic moment is approximately 0.4 $\mu_\text{B}$ per edge boron atom.
Similar edge-dependent behaviors are seen in other 2D materials. For example, armchair graphene nanoribbons are nonmagnetic semiconductors, while zigzag graphene nanoribbons exhibit edge spin polarization due to localized $\pi$ orbitals~\cite{Chakrabarty2018,Wakabayashi2001}. Similarly, armchair silicene nanoribbons are nonmagnetic and show a width-dependent band gap oscillation with a period of three, whereas zigzag silicene nanoribbons are antiferromagnetic semiconductors and become half-metallic under transverse electric fields~\cite{Ding2009,Song2010,Fang2013}.

\section{Tight-binding Hamiltonian for $\beta_{12}$-borophene nanoribbons}\label{A1}

\begin{figure*}
  \centering
\includegraphics[width=0.8\linewidth]{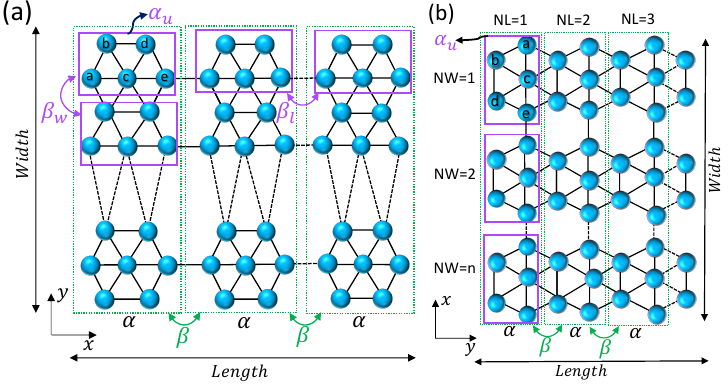}
\caption{Schematic illustration of armchair- and zigzag-edged $\beta_{12}$-BNR geometries. (a)~For the armchair-edged $\beta_{12}$-BNR, the fundamental five-atom unit cell ($\alpha_u$, highlighted by the purple rectangle) repeats $NW$ times along the ribbon width, forming a supercell ($\alpha$). The supercell is then periodically repeated along the ribbon length. Interaction matrices between adjacent fundamental cells along the width and length directions are denoted as $\beta_w$ and $\beta_l$, respectively, while interactions between neighboring supercells are represented by $\beta$. These matrices collectively define the tight-binding Hamiltonian. (b)~Corresponding schematic for the zigzag-edged $\beta_{12}$-BNR geometry, illustrating an alternative edge configuration.}
  \label{fBNR}
\end{figure*}
Figure~\ref{fBNR} illustrates the distinct nanoribbon geometries of $\beta_{12}$-borophene. In Fig.~\ref{fBNR}(a), the armchair-edged $\beta_{12}$-BNR is constructed by periodically repeating a fundamental five-atom unit cell, denoted as $\alpha_u$ (highlighted by the purple rectangle), along both the transverse and longitudinal directions. This repetition forms an extended supercell, $\alpha$, whose internal couplings are characterized by different interaction matrices: $\beta_w$ describes the hopping between adjacent $\alpha_u$ cells across the ribbon width, $\beta_l$ captures the interactions along the length, and $\beta$ accounts for the coupling between neighboring supercells. These matrices combine to yield a tight-binding Hamiltonian that encapsulates the essential physics of the system.
Fig.~\ref{fBNR}(b) shows the zigzag-edged $\beta_{12}$-BNR, where the same fundamental unit cell is employed but the edge termination introduces distinct boundary conditions. These differences affect the localization of edge states and modify the band dispersion, potentially leading to variations in transport properties and even topological characteristics. Overall, the interplay between atomic geometry, orbital contributions, and inter-cell couplings is central to understanding the electronic behavior of $\beta_{12}$-borophene nanoribbons.
To study the band structures of $\beta_{12}$-BNRs with N unit cells in the ribbon width, we first obtain the discrete Schrödinger equation corresponding to the Hamiltonian based on the $\beta_{12}$-borophene unit cell. The wave function phase shift for moving between adjacent unit cells is $e^{ika}$ (forward) or $e^{-ika}$ (backward), allowing the Hamiltonian calculation to account for the central scattering region interacting with the unit cell times the phase factor. Using Bloch's theorem, the total Hamiltonian for $\beta_{12}$-BNRs in momentum space is given by~\cite{Davoudiniya2021,Davoudiniya2021_imp}:

\begin{equation}
\mathcal{H} = \alpha + \beta^\dagger e^{-ika} + \beta e^{ika},
\end{equation}
where $\alpha$ represents the Hamiltonian of the central scattering region. The Hamiltonian matrix $\alpha_u$, which describes each unit cell, is given by:
\begin{equation}
\alpha_u \;=\;
\begin{pmatrix}
\varepsilon_a & t_{ab} & t_{ac} & 0 & 0 \\
t_{ab}    & \varepsilon_b & t_{bc} & t_{bd} & 0 \\
t_{ac}    & t_{bc}    & \varepsilon_c & t_{cd} & t_{ce} \\
0       & t_{bd}    & t_{cd}    & \varepsilon_d & t_{de} \\
0       & 0       & t_{ce}    & t_{de}    & \varepsilon_e
\end{pmatrix},
\end{equation}
$\beta_w$ describes interactions between neighboring unit cells across the ribbon width. Likewise, $\beta_l$ characterizes the interactions between adjoining unit cells along the transport direction. The explicit forms of these matrices are given by:
\begin{equation}
\beta_w \;=\;
\begin{pmatrix}
0 & 0 & 0 & 0 & 0 \\
0 & 0 & 0 & 0 & 0 \\
0 & 0 & 0 & 0 & 0 \\
0 & 0 & 0 & 0 & 0 \\
t_{ae} & 0 & 0 & 0 & 0
\end{pmatrix},
\beta_l \;=\;
\begin{pmatrix}
0 & 0 & 0 & 0 & 0 \\
t_{ab} & 0 & t_{bc} & 0 & 0 \\
0 & 0 & 0 & 0 & 0 \\
0 & 0 & t_{cd} & 0 & t_{de} \\
0 & 0 & 0 & 0 & 0
\end{pmatrix}.
\end{equation}

A super unit cell, $\alpha$, is defined as one column of unit cells across the ribbon width and one unit cell length in the transport direction. The interaction matrix $\beta$ between two super unit cells is:
\begin{equation}
\alpha = \begin{pmatrix}
\alpha_u & \beta_w & 0 & 0 \\
\beta_w^\prime & \alpha_u & \beta_w & 0 \\
0 & \beta_w^\prime & \alpha_u & \dots \\
0 & 0 & \dots & \ddots
\end{pmatrix}_{(N\times N)},
\beta = \begin{pmatrix}
\beta_l & 0 & 0 & 0 \\
0 & \beta_l & 0 & 0 \\
0 & 0 & \beta_l & \dots \\
0 & 0 & \dots & \ddots
\end{pmatrix}_{(N\times N)}
\end{equation}
For $N_T$ unit cells along the transport direction, the Hamiltonian matrix can be defined as:
\begin{equation}
\mathcal{H} = \begin{pmatrix}
\alpha & \beta & 0 & 0 \\
\beta^\prime & \alpha & \beta & 0 \\
0 & \beta^\prime & \alpha & \dots \\
0 & 0 & \dots & \ddots
\end{pmatrix}_{(N\times N_T)\times (N\times N_T)}
\label{Ham}
\end{equation}

\section{Quantum transport based on the Landauer-B\"{u}ttiker formula}
We analyze electronic transport through finite $\beta_{12}$-BNRs employing the Landauer-B\"uttiker approach, implemented through standard nonequilibrium Green’s function techniques. Within this formalism, the transport properties are determined from Green's functions describing the coupled system of a central scattering region and semi-infinite leads.

The surface Green's functions of the left ($g_L$) and right ($g_R$) electrodes are computed by iterative solutions as described in Refs.~\cite{Sancho_1985,PhysRevB.23.4997}:
\begin{subequations}
\begin{align}
g_L &= \left[ (\mathcal{E}+i\eta)I - \alpha - (\beta)^\dagger\bar{\tau} \right]^{-1}, \\
g_R &= \left[ (\mathcal{E}+i\eta)I - \alpha - \beta \tau \right]^{-1},
\end{align}
\end{subequations}
where the spin state $\sigma$ is labeled as spin-up ($\uparrow$) or spin-down ($\downarrow$) and $\mathcal{E}$ is the energy, $\eta$ is a small positive number ensuring numerical stability, $I$ is the identity matrix, and matrices $\alpha$ and $\beta$ describe the on-site and coupling terms in the leads, respectively. The transfer matrices $\tau$ and $\bar{\tau}$ are calculated iteratively:
\begin{subequations}
\begin{align}
\tau &= t_0 + \bar{t}_0 t_1 + \bar{t}_0 \bar{t}_1 t_2 + \cdots + \bar{t}_0 \bar{t}_1 \cdots t_n, \\
\bar{\tau} &= \bar{t}_0 + t_0 \bar{t}_1 + t_0 t_1 \bar{t}_2 + \cdots + t_0 t_1 \cdots \bar{t}_n,
\end{align}
\end{subequations}
with recursive relations given by:
\begin{subequations}
\begin{align}
t_n &= \left[I - t_{n-1} \bar{t}_{n-1} - \bar{t}_{n-1} t_{n-1}\right]^{-1}(t_{n-1})^2, \\
\bar{t}_n &= \left[I - t_{n-1} \bar{t}_{n-1} - \bar{t}_{n-1} t_{n-1}\right]^{-1}(\bar{t}_{n-1})^2,
\end{align}
\end{subequations}
and initial conditions:
\begin{subequations}
\begin{align}
t_0 &= \left[(\mathcal{E}+i\eta)I - \alpha\right]^{-1}(\beta)^\dagger, \\
\bar{t}_0 &= \left[(\mathcal{E}+i\eta)I - \alpha\right]^{-1}\beta,
\end{align}
\end{subequations}
Iterations continue until convergence of $t_n$ and $\bar{t}_n$ to below a predefined numerical tolerance $\delta$.

With the lead surface Green's functions obtained, the central-region Green's function $\mathcal{G}_C$ is computed via:
\begin{equation}
\mathcal{G}_C = \left[(\mathcal{E}+i\eta)I - \alpha - \Sigma_L - \Sigma_R\right]^{-1},
\end{equation}
where the self-energy matrices $\Sigma_L(\mathcal{E})$ and $\Sigma_R(\mathcal{E})$ describe the coupling between the central region and the left and right leads, respectively:
\begin{equation}
\Sigma_L = (\beta)^{\dagger} g_L \beta, \quad \Sigma_R = \beta g_R (\beta)^{\dagger}.
\end{equation}
The corresponding linewidth (coupling strength) functions for the left and right electrodes are:
\begin{equation}
\Gamma_{\{L,R\}} = i\,\bigl[\Sigma_{\{L,R\}} - (\Sigma_{\{L,R\}})^{\dagger}\bigr]
\end{equation}

The density of states, describing the available electronic states in the central region, is obtained as:
\begin{equation}
DOS = -\frac{1}{\pi} \,\mathrm{Im}\!\left[\mathrm{Tr}\,\mathcal{G}_C\right].
\label{dos}
\end{equation}

\section{Calculation of current with electron-phonon coupling}
\noindent
In this section, we provide additional details on electron--phonon interactions within the NEGF framework. The main text (Sec.~\ref{currentph}) outlines how the total current is split into an coherent part ($I_0$) and a phonon-assisted part ($I_{\mathrm{ph}}$). Here, we focus on the Holstein phonon self-energy, $\Sigma_{\mathrm{ph}}$, which enters the central-region Green's function and governs the inelastic processes that give rise to $I_{\mathrm{ph}}$.

\subsection{Phonon self-energy}
To incorporate e-ph interactions, we introduce the Holstein phonon self-energy. Transforming to the frequency domain, the lesser component of the phonon self-energy is given by
\begin{equation}
\Sigma_{ph}^{<}(\omega) \;=\; \frac{i}{2N}\sum_{\mathbf{q},\lambda}g_{\mathbf{q},\lambda}^{2}\int g_C^{<}(\mathcal{E})\,D^{<}(\omega - \mathcal{E})\,\frac{d\mathcal{E}}{2\pi},
\end{equation}
where the lesser phonon Green's function is
\begin{equation}
D^{<}(\omega) = -2\pi i\,n_B(\omega)\bigl[\delta(\omega - \omega_0) - \delta(\omega + \omega_0)\bigr]
\end{equation}
with $n_B(\omega)$ being the Bose-Einstein distribution function and $\omega_0$ the phonon frequency.

The lesser Green's function for the central region is assumed to be
\begin{equation}
g_C^{<}(\omega) = i\,f_C(\omega)\bigl[-2\,\mathrm{Im}\,g_C^r(\omega)\bigr].
\end{equation}
with the spectral function defined as
\begin{equation}
A_C(\omega) = -2\,\mathrm{Im}\,g_C^r(\omega).
\end{equation}
Substituting these expressions, the lesser phonon self-energy simplifies to
\begin{equation}
\Sigma_{ph}^{<}(\omega) \;=\; \frac{i}{2N}\sum_{\mathbf{q},\lambda}g_{\mathbf{q},\lambda}^{2}\,f_C(\omega)\,A_C(\omega)\,\bigl(2\,n_B(\omega_0)+1\bigr).
\end{equation}
Similarly, the greater phonon self-energy is given by
\begin{equation}
\Sigma_{ph}^{>}(\omega) \;=\; -\frac{i}{2N}\sum_{\mathbf{q},\lambda}g_{\mathbf{q},\lambda}^{2}\,\bigl[1 - f_C(\omega)\bigr]\,A_C(\omega)\,\bigl(2\,n_B(\omega_0)+1\bigr).
\end{equation}

\subsection{Green's functions and self-energies}
The total Green's function for the central region is obtained from:
\begin{equation}
G_C = \left[z - \mathcal{H}_C - \Sigma_C\right]^{-1}, \quad z = \epsilon + i\eta,,
\end{equation}
where $\mathcal{H}_C$ is the Hamiltonian of the central region and $\eta$ is a positive infinitesimal.
with the self-energy $\Sigma_C$ including contributions from the leads and phonons,
\begin{equation}
\Sigma_C = \Sigma_{\mathrm{leads}} + \Sigma_{\mathrm{phonons}} + \dots,
\end{equation}

For the left electrode, the Green's function satisfies
\begin{equation}
G_L = g_L + g_L\,\Sigma_L\,G_L,
\end{equation}
with the self-energy including both lead and phonon contributions:
\begin{equation}
\Sigma_L = \Sigma_L^r + \Sigma_{ph}^r + \dots,
\end{equation}
and one may also relate the coupling matrices via
\begin{equation}
V_L^\dagger\left[-2\,\mathrm{Im}\,g^r_L\right]V_L = \Gamma.
\end{equation}

}
\twocolumngrid

\bibliography{ref.bib}
\end{document}